\documentclass[pre,12pt,showpacs]{revtex4}
\usepackage{epsfig}
\usepackage{graphicx}

\normalsize

\begin{document}

\title{{\Large SUPERSTATISTICS OF BROWNIAN MOTION: A COMPARATIVE STUDY }}
\author{R. F. Rodr\'{\i}guez \footnote{%
Fellow of SNI, M\'{e}xico. Also at FENOMEC.} \footnote{%
Correspondence author. E-mail: zepeda@fisica.unam.mx. }}
\affiliation{Instituto de F\'{\i}sica, Universidad Nacional Aut\'{o}noma de M\'{e}xico.\\
Apdo. Postal 20-364, 01000, M\'{e}xico, D. F., M\'{e}xico. \\
and}
\author{I. Santamar\'{\i}a-Holek \footnote{%
Fellow of SNI, M\'exico. E-mail ivan@graef.fciencias.unam.mx }}
\affiliation{Facultad de Ciencias, Universidad Nacional Aut\'{o}noma de M\'{e}xico.\\
Circuito Exterior, Cd. Universitaria, 04510, M\'{e}xico, D. F., M\'{e}xico}

\begin{abstract}
The dynamics of temperature fluctuations of a gas of Brownian particles in
local equilibrium with a nonequilibrium heat bath, are described using an
approach  consistent with Boltzmann-Gibbs statistics ($BG$). We use
mesoscopic nonequilibrium thermodynamics ($MNET$) to derive a Fokker-Planck
equation for the probability distribution in phase space including the local
intensive variables fluctuations. We contract the description to obtain an
effective probability distribution ($EPD$) from which the mass density, van
Hove's function and the dynamic structure factor of the system are obtained.
The main result is to show that in the long time limit the $EPD$ exhibits a
similar behavior as the superstatistics distribution of nonextensive
statistical mechanics ($NESM$), therfore implying that the coarse-graining
procedure is responsible for the so called nonextensive effects.
\end{abstract}

\keywords{brownian motion - nonequilibrium temperature fluctuations -
multiplicative stochastic equations}
\pacs{ 05.70.Ln, 05.10.Gg, 05.40.-a, 87.10.+e}
\maketitle

\section{Introduction}

The existence of fluctuations of intensive parameters, such as the inverse
temperature, the chemical potential, the pressure or the energy dissipation
rate in a turbulent fluid, is a common feature in a variety of physical
systems in nonequilibrium states \cite{ausloos}, \cite{beck3}, \cite{beck1}, 
\cite{fluct-friction}. However, the theoretical description of these
fluctuations is well beyond the range of applicability of the theory of
fluctuations as developed by Onsager and Machlup \cite{onsager} and
subsequent generalization by Fox and Uhlenbeck \cite{fox}, and others \cite%
{ortiz}. This is essentially due to the fact that these theories are
restricted to describe the dynamics of fluctuations of extensive state
variables only.

One of the most used methods to deal with intensive variable fluctuations is
the superstatistics approach of $NESM$, \cite{tsallis}, \cite{abe}, 
\cite{beck2}, \cite{cohen1}. This theory claims that a proper description of
these fluctuations for systems with a sufficiently complex dynamics, demands
the introduction of a more general statistics --- Tsallis statistics ($TS$)
--- than the usual Boltzmann-Gibbs statistics ($BG$). Furthermore, according
to Refs. \cite{beck1}, \cite{cohen1}, $TS$ is only one of many more general
statistics --- superstatistics --- that can deal with the physical
situations mentioned above. An important and simple system where these
issues have been explored is Brownian motion in the presence of inverse
temperature fluctuations \cite{beck1}. In this reference the fluctuations of
the intensive parameter $1/T$ are not described as a fully time-dependent
phenomena, rather only their statistics is taken into account and postulated
without a physical justification, by assuming that if $\beta $ is a $\chi
^{2}$ distributed random variable. By averaging over its (static)
statistical distribution this model generates $TS$ for $v$ for a stationary
state. On this basis, it is suggested that $BG$ is no longer capable of
describing correctly this physical situation and that a new statistics ($TS$%
) is required.

The main motivation of the present work is to consider inverse temperature
fluctuations in the same model as in Ref. \cite{beck1}, but describing their
dynamics in terms of a stochastic process, and following their time
evolution towards equilibrium. By using $MNET$ \cite{mazur}, \cite{pnas}, 
\cite{reguera}, the time evolution of the nonequilibrium state is described
in the phase space of the particles, including the local intensive
fluctuations as state variables, by means of a Fokker-Planck equation ($FPE$%
). In the diffusion limit this description leads in a natural way to a
stochastic Smoluchowski equation for the contracted distribution function.
By following a method introduced by van Kampen to solve multiplicative
stochastic equations \cite{van kampen}, we derive an equation for the
effective distribution function ($EDF$), which still contains all the
effects produced by the induced (intensive) fluctuations. The long time
limit behavior of the $EDF$ as a function of position, is similar to the
behavior found in $NESM$ for the so-called superstatistical distribution. In
this way, our approach shows to be an alternative way to describe the
dynamics of intensive fluctuations consistently with $BG$.

We will proceed as follows. In Section 2, we present a derivation of the $%
FPE $, \cite{ivan1}. In Section 3 we model the temperature fluctuations by
means of a stochastic process and after contracting the $FPE$, we derive a
stochastic multiplicative Smoluchowski equation ($MSE$), valid for an
arbitrary stochastic dynamics of its coefficients. We conclude this section
by deriving from it\ a deterministic equation for the $EDF$. The solution of
this equation allows us to calculate the van Hove\'s function and the dynamic structure factor of the gas, which are obtained
in Section 4. In Section 5 we show that when the long time limit of the $EDF$
is evaluated as a function of the position, the contraction procedure and
the non-Gaussian corrections introduced by solving the $MSE$, yield a
behavior which reproduce the one proposed by $NESM$ for the same model. We
conclude with a discussion of the results of Section 5.

\section{Smoluchowski equation}

Following Ref. \cite{beck1}, we consider a driven nonequilibrium system
composed of regions (cells) of size $l$ where fluctuations of $1/T$ may
occur \cite{beck2}. In this model system, it is assumed that the local
temperature of a cell changes in a time scale, $\tau $, much larger than the
relaxation time a region needs to reach local equilibrium. Within each cell
a spherical test particle of radius $a$ and mass unity, performs Brownian
motion and then moves to another cell. Its velocity may be described by a
linear Langevin equation of the form%
\begin{equation}
\frac{dv}{dt}=-\gamma v+\sigma L(t),  \label{1}
\end{equation}%
where $\gamma >0$ is a friction constant, $L(t)$ is Gaussian white noise and 
$\sigma $ describes the strength of the noise. For this model the inverse
temperature $\beta =\gamma /\sigma ^{2}$ is no longer constant, but
fluctuates in space and time on the scales $l$ and $\tau $, respectively. As
a result, Brownian motion takes place on two time scales, one related to the
dissipation of the kinetic energy through the Stokes friction coefficient $%
\gamma =6\pi \eta a$, where $\eta $ is the shear viscosity of the host
fluid, and the second one is associated with the presence of temperature
fluctuations induced on the whole system by an external agent. These
fluctuations originate in the environment of the Brownian particles (heat
bath) and in some cases may therefore possess a complex dynamics.

In local equilibrium, the probability of finding a Brownian particle in
position $\vec{r}$ with instantaneous velocity $\vec{u}$ in the presence of
the temperature fluctuations $\delta T$, is described by the local
equilibrium probability density $P_{le}(\vec{r},\vec{u};\delta T)$. This
quantity may be determined through the calculation of the amount of work
exerted on a particle at temperature $T$ in the presence of an external
potential $\phi (\vec{r})$ and of a local temperature fluctuation $\delta
T(t)\equiv T-T_{B}$, where $T_{B}$ is the temperature of the bath \cite%
{mazur}, \cite{ivan1}, \cite{de groot}, \cite{landau sp}. If the spontaneous
thermal (equilibrium) fluctuations $\Delta T$ are such that $\Delta T\ll
\delta T$, the expression for $P_{le}$ is \cite{ivan1} 
\begin{equation}
P_{le}(u)=P_{0}e^{-\frac{m}{2k_{B}T}\left( u^{2}+\phi \right) }e^{-\frac{mc_{v}}{k_{B}}\frac{\left( \Delta T\right) ^{2}}{TT_{B}}},
\label{3}
\end{equation}%
where $m$ is the mass of the particle, $c_{v}\equiv \partial e/\partial
T_{B} $ is the specific heat at constant volume, $e$ is the internal energy
density and where the density fluctuations of the heat bath have been
neglected .

The dynamics of the Brownian particle may be described through the
nonequilibrium probability density $P(\vec{r},\vec{u},t;\delta T)$. The time
evolution equation for this quantity is derived from the entropy production $%
\partial \Delta s(t)/\partial t$, where the nonequilibrium entropy increment 
$\Delta s(t)$ is given by the generalized Gibbs entropy postulate%
\begin{equation}
\Delta s(t)=-k_{B}\int P\ln \frac{P}{P_{le}}d\vec{r}d\vec{u}.  \label{2}
\end{equation}
Since the probability conservation is expressed through the continuity
equation%
\begin{equation}
\frac{\partial P}{\partial t}=-\frac{\partial }{\partial \vec{r}}\cdot
\left( \vec{u}P\right) -\frac{\partial }{\partial \vec{u}}\cdot \left( \vec{V%
}_{\vec{u}}P\right) ,  \label{4}
\end{equation}%
\bigskip from Eqs. (\ref{3}), (\ref{2}) and (\ref{4}) and by using the
boundary conditions for the fluxes, we get an expression for the entropy
production where fluxes and forces may be clearly identified, \cite{pnas}, 
\cite{reguera}, \cite{ivan1}. By assuming linear relationships between
fluxes and forces we finally arrive at the Fokker-Planck equation 
\begin{equation}
\frac{\partial }{\partial t}P+\frac{\partial }{\partial \vec{r}}\cdot \left( 
\vec{u}P\right) =\frac{\partial }{\partial \vec{u}}\cdot \left[ \left( \beta 
\vec{u}+\frac{\partial \phi }{\partial \vec{r}}\right) P+\frac{k_{B}T}{m}%
\beta \frac{\partial P}{\partial \vec{u}}\right] ,  \label{5}
\end{equation}%
where we have taken the coupling coefficient for the external force ${%
\partial \phi }/{\partial \vec{r}}$ as one, in accordance with the
phenomenological description we are using \cite{nonmarkov}, \cite{adelman}.
It should be pointed out that in arriving at Eq. (\ref{5}) we have neglected
the contribution arising from the usual thermal fluctuations in Eq. (\ref{3}%
), in accordance with the assumption $\Delta T\ll \delta T$.\ 

The Smoluchowski equation for the reduced probability distribution $\rho (%
\vec{r},t)\equiv \int Pd\vec{u}$ can be derived in the limit of long times,
by calculating the evolution equations for the first three moments of $P$
over $\vec{u}$-space, namely, the mass density $\rho (\vec{r},t)\equiv \int
Pd\vec{u}$, the momentum density $\rho \vec{v}(\vec{r},t)$ and the stress
tensor density. For sufficiently long times, one may assume that the
stresses and momenta have relaxed, and by combining them we obtain a
constitutive relation for the diffusion flow (see Ref. \cite{ivan1} for
details). After substitution of this constitutive relation into the
continuity equation we are lead to 
\begin{equation}
\frac{\partial \rho }{\partial t}=\beta ^{-1}\frac{\partial }{\partial \vec{r%
}}\cdot \left( \rho \frac{\partial \phi }{\partial \vec{r}}\right) +D\frac{%
\partial ^{2}\rho }{\partial \vec{r}^{2}},  \label{6}
\end{equation}%
where in the last term we have taken into account that $T=T_{B}+\delta T$
and Eq. (\ref{6}) contains the diffusion coefficient%
\begin{equation}
D\equiv \left( {k_{B}T_{B}}/{m\beta }\right) \left( 1+{\delta T}/{T_{B}}%
\right) ,  \label{7}
\end{equation}%
which explicitly incorporates the effects of the externally induced
fluctuations on the temperature of the bath. It is convenient to stress that
equations similar to (\ref{6}) have been previously derived in the context
of slow relaxation systems such as supercooled colloidal suspensions and
granular systems, among others \cite{slow}, \cite{jpcm}. However, in these
cases, the temperature $T$ was assumed to be a decaying function of time,
related with an activated process which controls the relaxation. In
contrast, in the present work we considered the more general situation where
the temperature fluctuation $\delta T(t)$ is assumed to be a stochastic
process, which converts Eq. (\ref{6}) into a multiplicative stochastic
partial differential equation ($MSE$).

\section{Effective probability distribution}

With the purpose of obtaining macroscopic local variables which eventually
may be measured, such as the dynamic structure factor, the van Hove's
function, or the local mass density, we contract the description over
velocities and $\delta T(t)$. To this end we define the $EDF$ as 
\begin{equation}
\langle \rho (\vec{r},t)\rangle \equiv \int P(\vec{r},\vec{u},t;\delta T)d%
\vec{u}d(\delta T).  \label{7a}
\end{equation}%
A time evolution equation for the $EDF$ is obtained by deriving an
approximate solution of Eq. (\ref{6}) through the use of a method proposed
by van Kampen which yields an equation for the average $\langle \rho (\vec{r}%
,t)\rangle $, \cite{van kampen}. This approximation takes the form of a
series expansion in powers of the Kubo number, $\alpha \tau _{c}$, which is
assumed to be small. Here $\alpha $ is a parameter measuring the magnitude
of the fluctuations in the coefficients of Eq. (\ref{6}) and $\tau _{c}$
denotes their finite autocorrelation time \cite{van kampen}.

This derivation may be easily generalized to consider the case where the
particle is also under the influence of a harmonic force derived form the
potential $\phi =\omega _{0}^{2}\vec{r}^{2}/2$, where $\omega _{0}$ denotes
its characteristic frequency. Let us assume that $\delta T(t)$ is of the form%
\begin{equation}
\delta T(t)/T_{B}=\alpha \zeta (t),  \label{8}
\end{equation}%
with $\zeta (t)$ is an arbitrary stochastic process with zero mean and
autocorrelation%
\begin{equation}
\langle \zeta (t-\tau )\zeta (t)\rangle =\Pi (\tau ).  \label{9}
\end{equation}%
Accordingly, we recast Eq. (\ref{6}) as the following multiplicative
stochastic equation for $\rho (\vec{x},t)$, 
\begin{equation}
\frac{\partial \rho }{\partial \tilde{t}}=\tilde{\nabla}\cdot (\rho \vec{x}%
)+\Gamma \tilde{\nabla}^{2}\rho +\alpha \Gamma \zeta (\tilde{t})\tilde{\nabla%
}^{2}\rho .  \label{10}
\end{equation}%
where for convenience we have introduced the dimensionless variables $\vec{x}%
\equiv a^{-1}\vec{r}$, $\tilde{t}\equiv \omega ^{2}\beta ^{-1}t$ and the
operator $\tilde{\nabla}$ indicates differentiation with respect to $\vec{x}$%
. The parameter $\Gamma \equiv k_{B}T_{B}/ma^{2}\omega ^{2}$ is the ratio
between the thermal energy of the bath and the energy supplied by the
external source, while maintaining the nonequilibrium state. By taking the
Fourier transform over the space variables, Eq. (\ref{10}) is rewritten as

\begin{equation}
\frac{\partial }{\partial \tilde{t}}\hat{\rho}(\vec{q},t)=-\vec{q}\cdot 
\frac{\partial \hat{\rho}}{\partial \vec{q}}-\Gamma \left( q^{2}+\alpha
\zeta (\tilde{t})q^{2}\right) \hat{\rho},  \label{11}
\end{equation}%
where the caret stands for the Fourier transform of a quantity.

To implement van Kampen's method explicitly, it is convenient to identify
the systematic and stochastic operators on the right hand side ($r.h.s.$) of
Eq. (\ref{11}) \ as 
\begin{equation}
A_{0}\equiv -\vec{q}\cdot \frac{\partial }{\partial \vec{q}}-\Gamma q^{2},
\label{12}
\end{equation}%
and 
\begin{equation}
A_{1}(t)\equiv -\alpha \Gamma \zeta (t)q^{2}.  \label{13}
\end{equation}%
Following van Kampen's method for times satisfying $\tau _{c}\ll t\ll \alpha
^{-1}$, the average $\langle \hat{\rho}(\vec{q},t)\rangle $ over the
realizations of $\zeta (t)$ obeys by itself the approximate non-stochastic
equation 
\begin{equation}
\frac{\partial \langle \hat{\rho}\rangle }{\partial \tilde{t}}=\left[
A_{0}+\alpha ^{2}\int_{0}^{\infty }d\tau \langle A_{1}(\tilde{t})e^{\tau
A_{0}}A_{1}(\tilde{t}-\tau )e^{-\tau A_{0}}\rangle \right] \langle \hat{\rho}%
\rangle .  \label{14}
\end{equation}%
After evaluating the action of the operators $A_{0}$ and $A_{1}$, Eq. (\ref%
{14}) may be rewritten in the more compact form 
\begin{equation}
\frac{\partial \langle \hat{\rho}\rangle }{\partial \tilde{t}}=-\left( \vec{q%
}\cdot \frac{\partial }{\partial \vec{q}}+q^{2}\Gamma \right) \langle \hat{%
\rho}\rangle +\Gamma ^{2}\alpha ^{2}C(\tilde{t})q^{4}\langle \hat{\rho}%
\rangle ,  \label{15}
\end{equation}%
where $C(\tilde{t})$ is defined in terms of the autocorrelation of the noise 
\begin{equation}
C(\tilde{t})\equiv \int_{0}^{\infty }d\tau \langle \zeta (\tilde{t})\zeta (%
\tilde{t}-\tau )\rangle ,  \label{16}
\end{equation}%
which is so far arbitrary. Note that the second term on the $r.h.s.$ of Eq. (%
\ref{15}) incorporates in an average way the effects of the temperature
fluctuations on the dynamics of the system. It also introduces time
dependent non-Gaussian corrections and gives rise to non-stationary effects
on the dynamics.

If for simplicity in the discussion we consider the special case where the
temperature fluctuations may be represented by a stationary stochastic
process, $C(t)=C(0)\equiv C_{0}$, and the steady state solution of Eq. (\ref%
{15}) reduces to 
\begin{equation}
\langle \hat{\rho}(q)\rangle =\rho _{0}\,e^{-\frac{\Gamma }{2}q^{2}}\,e^{%
\frac{\alpha ^{2}\Gamma ^{2}C_{0}}{4}q^{4}},  \label{17a}
\end{equation}%
where $\rho _{0}$ is a normalization factor. To comply with the
approximation of van Kampen's method, this solution is expanded up to order $%
\alpha ^{2}$ and then the inverse Fourier transform of the resulting
expression is taken. This leads to a solution which contains explicit
non-Gaussian corrections to the steady state distribution function, 
\begin{equation}
\langle \rho (x)\rangle =\frac{\rho _{0}}{\sqrt{2\pi \Gamma }}e^{-\frac{1}{%
2\Gamma }x^{2}}\left[ 1+\frac{1}{4}\alpha ^{2}\Gamma ^{2}C_{0}f(x)\right] ,
\label{18a}
\end{equation}%
with $f(x)=\Gamma ^{-2}x^{4}-6\Gamma ^{-1}x^{2}+3$. For future convenience
we expand the above expression in a Taylor series in $x$, 
\begin{equation}
\langle \rho (x)\rangle _{s}\simeq \frac{\rho _{0}(2\pi \Gamma )^{-1/2}}{1+%
\frac{x^{2}}{2\Gamma }+...}\left[ 1+\frac{1}{4}\alpha ^{2}C_{0}f(x)\right] .
\label{19a}
\end{equation}

Notice that a different (but similar) expansion can be carried out in terms
of the well known formula $e^{-y}\simeq \left( 1-\frac{y}{b}\right) ^{b}$,
which leads to%
\begin{equation}
\langle \rho (x)\rangle_{e} \simeq \rho _{0}(2\pi \Gamma )^{-1/2}\left( 1-%
\frac{x^{2}}{2\Gamma b}\right) ^{b}\left[ 1+\frac{1}{4}\alpha ^{2}C_{0}f(x)%
\right].  \label{19b}
\end{equation}

\section{van Hove's function and structure factor}

Owing to its direct relation with experimental techniques, we shall now
calculate the density correlation function \cite{boon}. The space Fourier
transform of this quantity is the van Hove function $\hat{F}$, 
\begin{equation}
\hat{F}(\vec{q},\tilde{t})\equiv \langle \rho (\vec{q},\tilde{t})\rho (\vec{q%
}_{0},\tilde{t}_{0})\rangle ,  \label{17}
\end{equation}%
where $\vec{q}_{0}$ will be considered as a reference wave vector, $\tilde{t}%
_{0}$ is an initial time and, as before, the bracket $\langle ..\,\rangle $
denotes an average over the realizations of the external noise $\zeta (t)$.
\ \ 

The time evolution equation of this quantity is obtained by multiplying Eq. (%
\ref{11}) by $\hat{\rho}(\vec{q}_{0},\tilde{t}_{0})$ and using Eqs. (\ref{12}%
-\ref{13}), yielding 
\begin{equation}
\frac{\partial }{\partial \tilde{t}}\hat{F}^{\ast }(\vec{q},t)=\left[
A_{0}+A_{1}(\tilde{t})\right] \hat{F}^{\ast }(\vec{q},\tilde{t}),  \label{18}
\end{equation}%
where we have used the notation $\hat{F}^{\ast }(\vec{q},t)\equiv
\left\langle \rho (\vec{q},t)\rho (\vec{q}_{0},t_{0})\right\rangle $. Since
Eq. (\ref{18}) is of the same type as Eq. (\ref{11}), van Kampen's method
also leads to the evolution equation 
\begin{equation}
\frac{\partial \hat{F}}{\partial \tilde{t}}=-\left( \vec{q}\cdot \frac{%
\partial }{\partial \vec{q}}+q^{2}\Gamma \right) \hat{F}+\Gamma ^{2}\alpha
^{2}C(\tilde{t})q^{4}\hat{F},  \label{19}
\end{equation}%
for $\hat{F}(\vec{q},\tilde{t})$ defined by Eq. (\ref{17}). Again, for the
special case where the temperature fluctuations may be represented by a
stationary stochastic process, the steady state solution of Eq. (\ref{19})
has the form (\ref{17a}),

\begin{equation}
\hat{F}_{s}(\vec{q})=e^{-\frac{\Gamma }{2}q^{2}}\,e^{\frac{\alpha ^{2}\Gamma
^{2}C_{0}}{4}q^{4}}.  \label{20}
\end{equation}%
Furthermore, using this result and the fact that Eq. (\ref{19}) is a linear
hyperbolic partial differential equation which can be solved by using the
method of the characteristics, we arrive at the following time dependent
solution of Eq. (\ref{19})

\begin{equation}
\hat{F}(\vec{q},t)\simeq \left( 1+\frac{\alpha ^{2}\Gamma ^{2}C_{0}}{4}%
q^{4}\right) e^{-\frac{\Gamma }{2}q^{2}-z^{2}(q,t)}.  \label{21}
\end{equation}%
Again, we have only retained terms up to order $\alpha ^{2}$ to be
consistent with the approximation used in van Kampen's method, and we have
used the scaling variable $z(q,t)\equiv t-\ln |q|$. Since the $z$ dependence
of the solution of Eq. (\ref{19}) should decay with $z$, we have chosen the
form $e^{-z^{2}}$. Equation (\ref{21}) actually imply that $F(\vec{q},t)$
has a similar behavior as the probability density $\left\langle \rho (\vec{q}%
,t)\right\rangle $. Figures 1a and 1b show a plot of $\hat{F}(\vec{q},t)$ as
a function of $\vec{q}$ and $t$ for fixed values of $t$ and $\vec{q}$,
respectively. They show that the presence of temperature fluctuations indeed
modifies the behavior of $F(\vec{q},t)$ with respect to the case where no
intensive fluctuations are present. We have expressed the wave vector as $%
\vec{q}\equiv q\widehat{k}$, where $\widehat{k}$ is a unit vector in the
direction of $\vec{q}$. It is important to point out that, in order to
represent Eq. (\ref{21}) by Figs. 1 (and Eq. (\ref{23}) through Figs. 2), we
have chosen values for the coefficients $\Gamma $, $\alpha $ and $C_{0}$
consistently with the approximations introduced in the previous section.
This means that the physical parameters ($T$, $\omega $, $\beta $)
characterizing the system are determined.
\begin{figure}[tbp]
{}
\par
\centering
\includegraphics*[width=6cm]{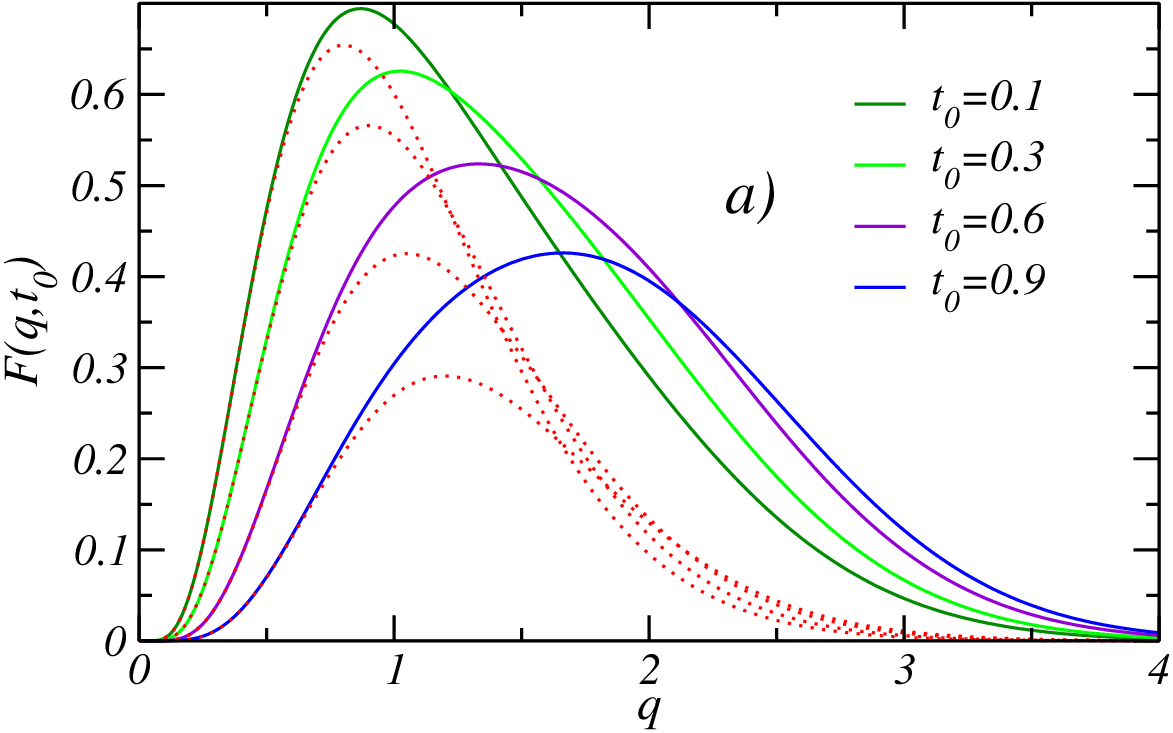}\hspace{0.4cm}
\includegraphics*[width=6cm]{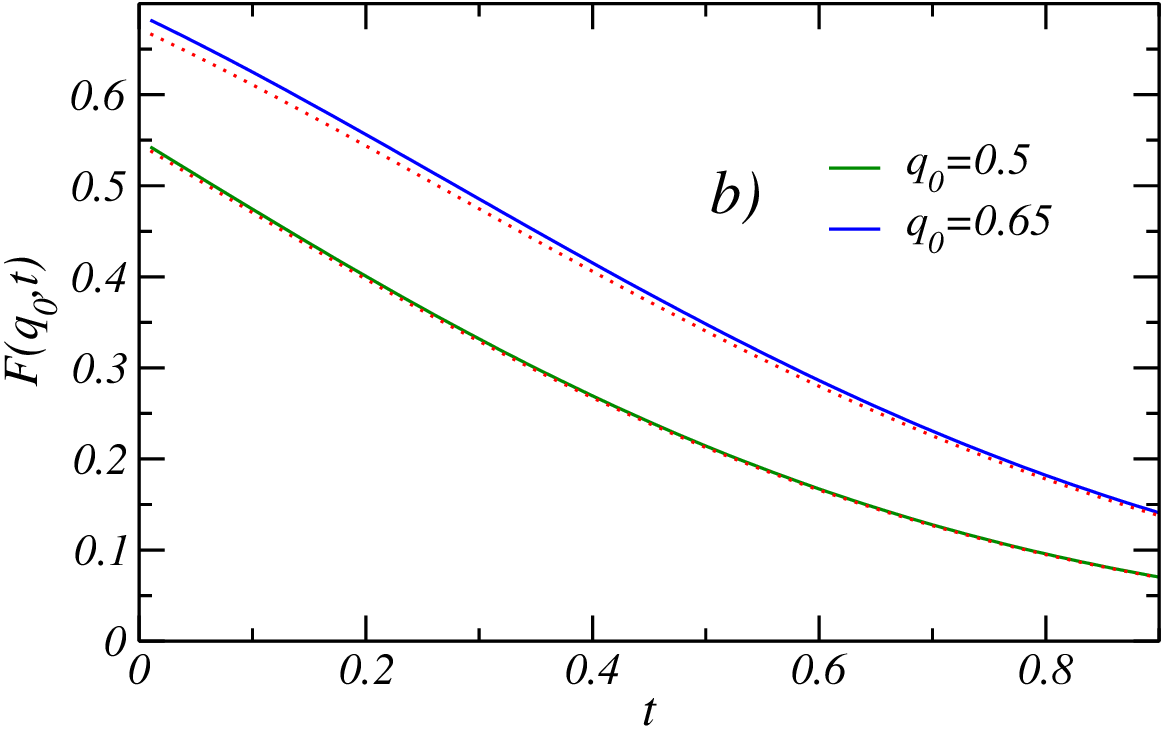}
\par
{\footnotesize {\ } \vspace{.0cm} }
\caption{{\emph{a})} $\widehat{F}(\vec{q},t)$ vs. $q$. The red lines
correspond to the case without temperature fluctuations. The remaining lines
green to blue, are obtained from Eq. (\ref{21}) for $t=0.1$, $0.3$, $%
0.6$, $0.9$, respectively. {\emph{b})} $F(\vec{q},t)$ vs. $t$ for fixed
values of $q=0.5$ (green) and $q=0.6$ (blue). The value of the parameters
are $C_{0}=0.8$, $\Gamma =1$ and $\protect\alpha =0.8$. }
\label{Fig 1a}
\end{figure}

Fig. 1a shows that the external fluctuations induce significant corrections
to $F(\vec{q},t)$ for values of $q>0.5$ for various fixed time values (see
caption). The red lines correspond to the case without (intensive)
temperature fluctuations, whereas the green to blue lines correspond to the
case when these fluctuations are present. This figure implies that intensive
temperature fluctuations slow down the decay of the density-density
correlation function. On the other hand, Fig. 1b shows the corrections
corresponding to the time dependence of van Hove's function for $q=0.5$
(green) and $q=0.6$ (blue).

The Fourier's transform of $\hat{F}(\vec{q},t)$ with respect to time defines
the dynamic structure factor $\hat{S}(q,\omega )$, 
\begin{equation}
\tilde{S}(q,\omega )=\frac{\hat{S}(q,\omega )}{S(q_{0},0)}=e^{i\omega \ln
|q|-\frac{\Gamma }{2}q^{2}(1-\frac{\Gamma }{2}\alpha ^{2}C_{0}q^{2})}.
\label{22}
\end{equation}%
Expanding this function up to first order in its argument, we arrive at 
\begin{equation}
\tilde{S}(q,\omega )=G^{-1}(q,\omega )\left[ 1+\frac{\Gamma }{2}q^{2}(1-%
\frac{\Gamma }{2}\alpha ^{2}C_{0}q^{2})\right] ,  \label{23}
\end{equation}%
where we have defined the inverse propagator 
\begin{equation}
G^{-1}(q,\omega )\equiv \left[ 1+\frac{\Gamma }{2}q^{2}(1-\frac{\Gamma }{2}%
\alpha ^{2}C_{0}q^{2})\right] ^{2}+\left[ \omega \ln |q^{2}|\right] ^{2}.
\label{24}
\end{equation}%
\begin{figure}[tbp]
{}
\par
\centering
\includegraphics*[width=6cm]{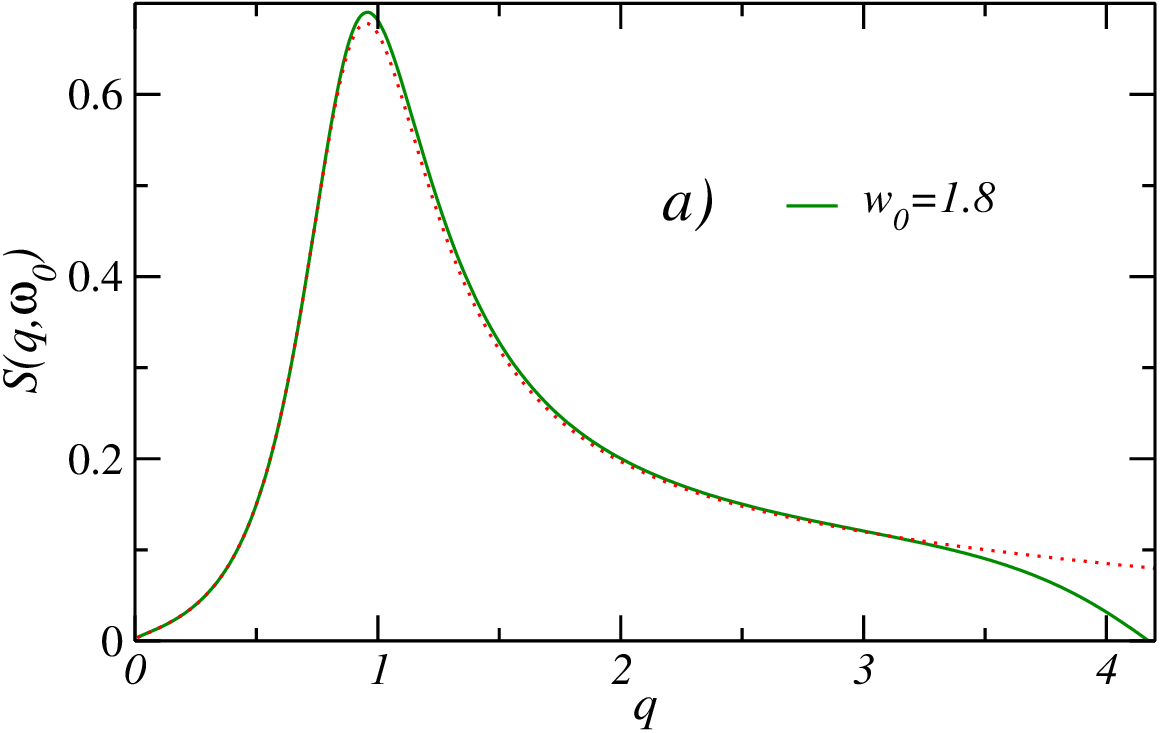}\hspace{0.4cm}
\includegraphics*[width=6cm]{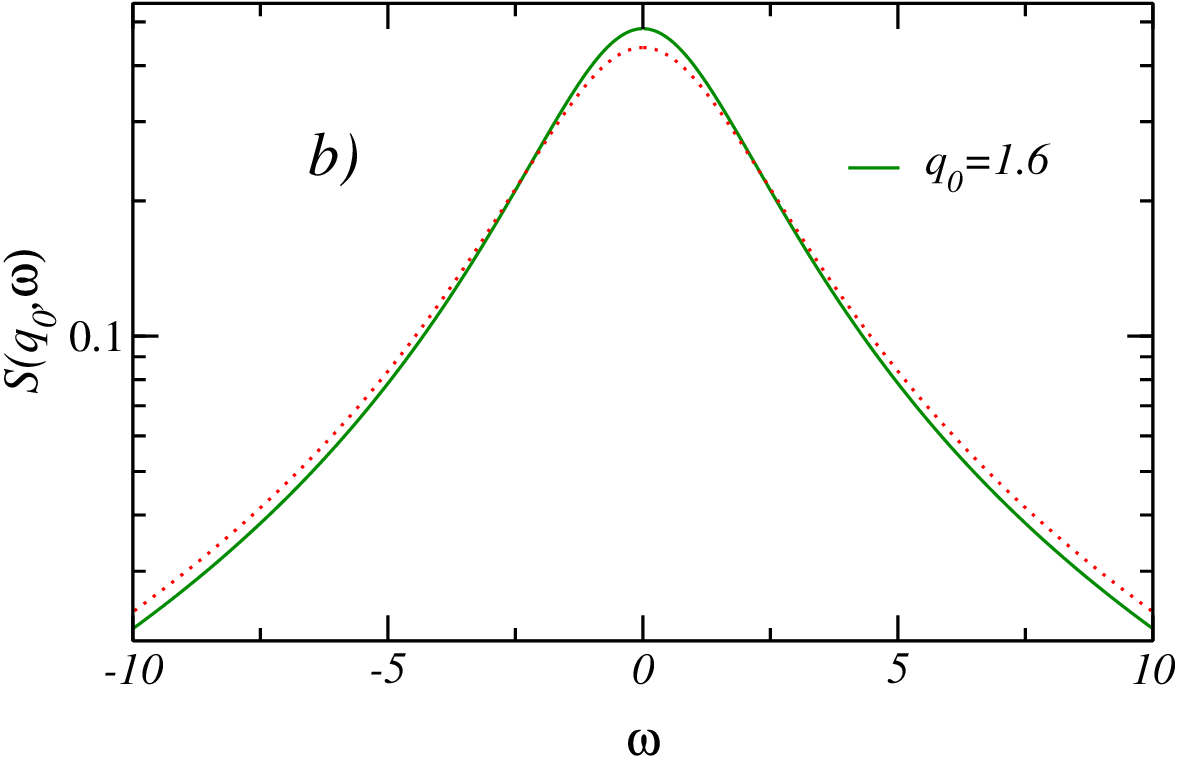}
\par
{\footnotesize {\ } \vspace{.0cm} }
\caption{{\emph{a})} $S(\vec{q},\omega )$ vs. $q $ for $
\omega=1.8$ and {\emph{b})} $S(\vec{q},\omega )$ vs. $q $ for $q=1.6$%
. The red lines correspond to the case without temperature fluctuations
whereas the green line include temperature fluctuations for $\alpha %
=0.8$, $C_{0}=0.8$ and $\Gamma =1$. }
\label{RI67}
\end{figure}
Eq. (\ref{23}) explicitly shows that the presence of the intensive
fluctuations indeed modify the behavior of the structure factor, as
expected. The presence of powers of $q$ higher than $q^{2}$ suggests the
existence of long-range correlations induced by intensive temperature
fluctuations. In Figs. 2 we show the projections of the structure factor for
fixed $q$ (Fig. 2a) and for fixed $\omega $ (Fig. 2b). Note that in Fig. 2b
the green line is $\sim 1.2$-times larger than the red one for $\omega =0$.
If $\omega \sim 2.5$ the green line has the same value as the red one and
for $\omega \simeq 10$ the green line is $\sim 0.6$-times larger than the
red one. Clearly, the presence of temperature fluctuations induces an
anomalous behavior of the structure factor in the $\omega $-fixed plane. It
should be pointed out that in this case, our results are only valid for $%
q<(c\alpha ^{2}\Gamma )^{-1/2}(1+\sqrt{1+4c\alpha ^{2}})$.

\section{Comparison with Superstatistics}

In order to compare the results obtained in this work with those derived
from the superstatistics approach reported in Refs. \cite{beck2,beck3}, we
recall that in these references it is argued that the stochastic
differential equation (\ref{1}) with a $\chi ^{2}$ distributed $\beta $,
gives rise to the generalized canonical distributions of $NESM$ \cite%
{tsallis},%
\begin{equation}
\rho _{NESM}(x)=\left[ 1+\frac{\beta _{0}}{n}x^{2}\right] ^{-\frac{n+1}{2}},
\label{25}
\end{equation}%
where $n$ is a positive integer and $\beta _{0}=\left\langle \beta
\right\rangle $. Eq. (\ref{25}) contains two free parameters ($\beta _{0}$
and $n$) that must be fitted in order to represent the function. Both
parameters arise from the initial assumption that the inverse temperature is
a $\chi ^{2}$ distributed variable; however, the physical meaning of these
parameters remains unknown. Moreover, it is interesting to notice that the
form of Eq. (\ref{25}) is independent of the nature of the variable of
interest. For example, in Ref. \cite{beck2}, a similar expression (with one
more parameter) is used in order to fit the distribution of velocities in a
fluid undergoing stationary turbulence. 
\begin{figure}[tbp]
{}
\par
\centering \mbox{\resizebox*{8cm}{!}{\includegraphics{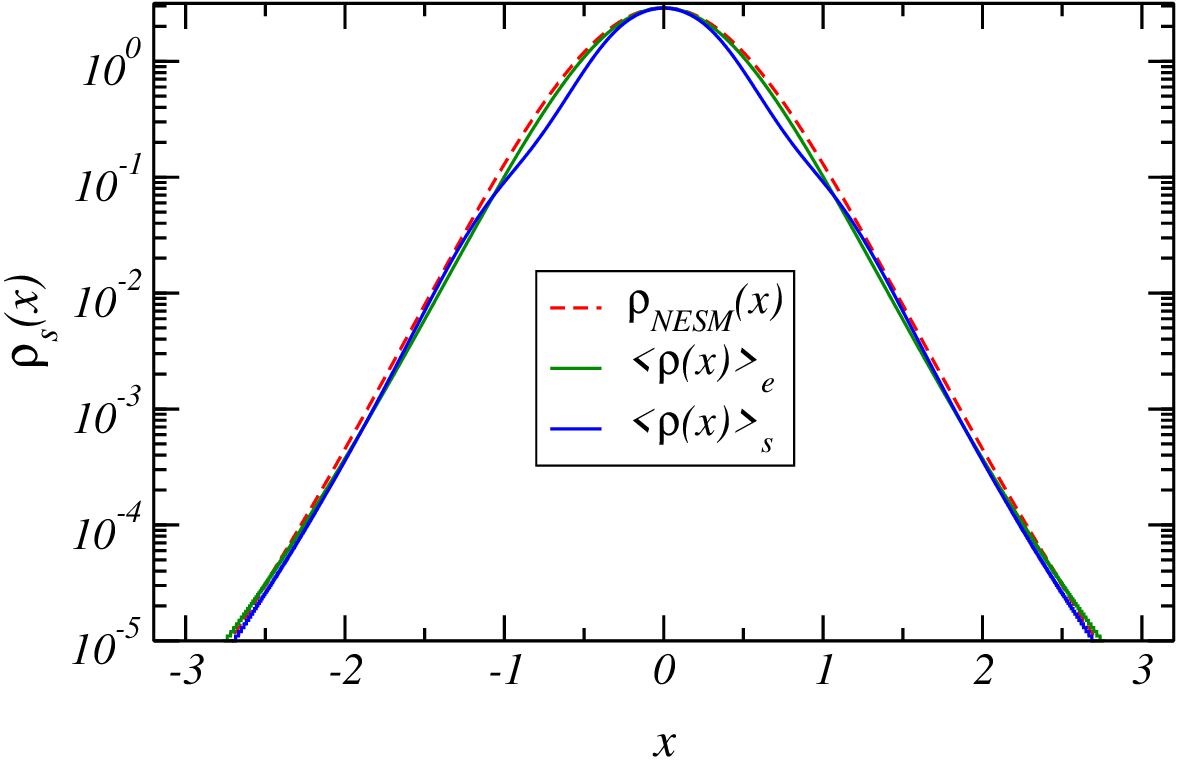}}}
\par
{\footnotesize {\ } \vspace{.0cm} }
\caption{The blue line represents the stationary $EDF$ as given by Eq. (%
\protect\ref{19a}) vs. $x$, for $n=18$ and with $\protect\alpha =0.5$, $%
C_{0}=0.25$, $\Gamma =0.13$. The green line represents an expansion of \ the
exponential in the stationary $EDF$ by using Eq. (\protect\ref{19b}) with $%
b=8$ and $\protect\alpha =0.02$, $C_{0}=0.12$ and $\Gamma =0.12$. }
\label{velpromedio}
\end{figure}

Figure 3 shows a comparison between our steady state distribution function,
as given by Eq. (\ref{19a}) (blue line), and the above expression for $\rho
_{NESM}(x)$ (red line), for different values of the parameter $n$. The green
line corresponds to Eq. (\ref{19b}). The difference between these curves
lies in the fact that the superstatistical method \cite{beck2}, essentially
analyzes the statistics of a system in local equilibrium, and considers the
fluctuations of the intensive parameters to be of a static nature. On the
contrary, Eq. (\ref{17a}) represents the long time limit of the dynamic
analysis that has taken into account the (irreversible) time dependent
process associated with the intensive fluctuations. Moreover, the parameters
entering into Eq. (\ref{19a}) are determined by the nature of the system and
must satisfy the conditions imposed by the approximations of Sec. III, ($%
\tau _{c}\alpha \ll 1$). In order to find a good comparison between the red
and blue lines, we have chosen the adequate order of the polynomial arising
from the Taylor expansion of the exponential.

It should be pointed out that in all cases the value of $\Gamma $ controls
the form of the peak, whereas the strength of the noise $\alpha $ and the
noise correlation $C_{0}$ control the tail of the distribution. The larger
the values of $\alpha $ and $C_{0}$, the larger the inflexion intermediate
values of $x$.

\bigskip

\section{Discussion}

In this work we have analyzed the problem of Brownian motion introduced in
Ref. \cite{beck2} by using a mesoscopic approach ($MNET)$. The time
evolution of the nonequilibrium state of the system is described by means of
a Fokker-Planck equation for the probability distribution in the phase space
of the particles including local intensive parameter fluctuations. The basic
difference with the point of view adopted in \cite{beck2} is that the
dynamics of the (intensive) temperature fluctuations is explicitly taken
into account by representing them as a stochastic process.

An essential element in our approach is the contraction of the above
description which yields a stochastic multiplicative Smoluchowski equation
for the (local) reduced probability density. The (approximate) solution of
this equation yields a deterministic equation for an effective probability
distribution ($EPD$) valid for an arbitrary stochastic dynamics of the
random coefficients. As a consequence, we calculated two measurable
quantities, the van Hove's function and the dynamic structure factor of the
system; however, we are not aware of any experimental results to compare
with these predictions of our model.

The most important conclusion that arises from our approach is that the long
time limit of our $EPD$ as a function of position, exhibits a similar
behavior to the so called superstatistical distribution in nonextensive
statistical mechanics ($NESM$), \cite{beck2}. This result shows that the
dynamics of intensive parameter fluctuations is fully consistent with
Boltzmann-Gibbs statistics. The coarse-graining procedure and the time
dependent non-Gaussian corrections introduced in solving the $MSE$, gave
rise to non-stationary effects on the dynamics and generated those features
of the $EPD$ which are usually identified with the so called nonextensive
effects of $NESM$.

Our approach can be used for both, stationary and non-stationary
fluctuations of the intensive parameters. However, it should be stressed
that our approach and conclusions are only applicable to the model analyzed
here and to the class of nonequilibrium states we have considered in the
present work. If for other nonequilibrium systems with a complex dynamics
similar conclusions can be drawn, is an open issue that remains to be
assessed.

\textbf{Acknowledgments}

Useful discussions and comments from Prof. J. M. Rub\'i are gratefully appreciated. Financial support from grant UNAM-DGAPA IN-108006 is acknowledged.

\end{document}